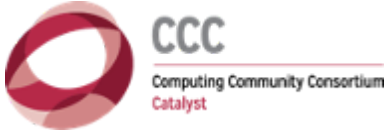

# Advancing Computing's Foundation of US Industry & Society
*A Computing Community Consortium (CCC) Quadrennial Paper*

*Thomas M. Conte (Georgia Institute of Technology), Ian T. Foster (Argonne National Laboratory), William Gropp (University of Illinois at Urbana-Champaign) and Mark D. Hill (University of Wisconsin-Madison)*

While past information technology (IT) advances have transformed society, future advances hold even greater promise. For example, we have only just begun to reap the changes from artificial intelligence (AI), especially machine learning (ML). Underlying IT's impact are the dramatic improvements in computer hardware, which deliver performance that unlock new capabilities. For example, recent successes in AI/ML required the synergy of improved algorithms and hardware architectures (e.g., general-purpose graphics processing units). However, unlike in the 20th Century and early 2000s, tomorrow's performance aspirations must be achieved without continued semiconductor scaling formerly provided by Moore's Law and Dennard Scaling. How will one deliver the next 100x improvement in capability at similar or less cost to enable great value? Can we make the next AI leap without 100x better hardware?

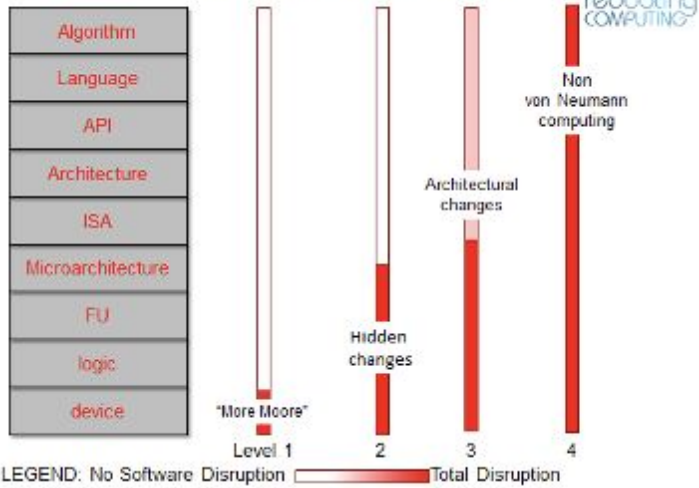

This whitepaper argues for a multipronged effort to develop new computing approaches beyond Moore's Law to advance the foundation that computing provides to US industry, education, medicine, science, and government. This impact extends far beyond the IT industry itself, as IT is now central for providing *value* across society, for example in semi-autonomous vehicles, tele-education, health wearables, viral analysis, and efficient administration. Herein we draw upon considerable visioning work by CRA's Computing Community Consortium (CCC) and the IEEE Rebooting Computing Initiative (IEEE RCI), enabled by thought leader input from industry, academia, and the US government.

In the 20th Century, Moore's Law's rapid doubling of transistors per chip (for similar cost) facilitated an era, now ending, where software and applications could be developed *independently* from computer hardware, because hardware maintained a fixed interface to software (instruction set architecture). The good news is that there are several ways to restart the meteoric rise of computer performance. The bad news is that the more revolutionary of these approaches will require not only significant hardware investment but also significant software rewriting and software-hardware "co-design". This disruption will entail significant technological risk that can exceed the portfolios of individual companies and industries. Thus, pre-competitive government investment is critical to enable future success.

We structure the rest of this whitepaper using the figure above. On the left of this figure, we see layers of computing technology (the "computing stack), starting with a computer program's algorithm and its programming language choice and extending down through the architecture and ultimately to the devices that comprise the computer's hardware. On the right of the figure, we show four levels of approach, numbered from 1 to 4, to "rebooting computing." Each level involves progressively greater disruption of today's technologies while also promising progressively greater opportunities. Some of the ideas that we consider here are relatively new, while others have previously been considered, but deserve revisiting now that Moore's law no longer dominates over other approaches.

**Level 1: "More Moore:"** This approach aims to extend Moore's law past the year 2025 somehow. Although the least disruptive to today's computing stack, the challenges inherent in this approach are great and potentially insurmountable. The basis for computing since the early 1980s is the CMOS (complementary metal oxide semiconductor) switch that we now scale near its 2D limits. One promising approach is to go 3D, although 3D chips introduce many challenges in manufacturing, cooling, and likely will require Level 2-3 changes to use well. Longer-term work should also examine new, more energy efficient and faster kinds of switch, although any such switch will ultimately also be limited by physical laws. These advances will be difficult to achieve for US society without unencumbered access to chip manufacturing as a foundation.

**Level 2: Hidden changes:** There is potential for using novel ways to construct computer internals while only disrupting part of the computing stack, thus mitigating the impact on software and applications. These techniques include adiabatic/reversible and cryogenic/superconducting computing. Adiabatic computing exploits the phenomenon that power in a computer circuit is consumed only when the number of inputs is reduced to a smaller number of outputs. Recycling unused inputs can save significant power, but requires different devices than CMOS. Superconducting computing aims to exploit different devices than the semiconductor industry produces today which when connected in new ways and cooled to very low temperatures (e.g., -452˚F where electrons can travel with zero resistance) can run at much higher frequencies than today's computers. If today's example superconducting systems can be moved from the lab into practice, there is a potential for new computers that still run today's software base, as well as providing a technological interface to quantum computers.

**Level 3: Architectural changes:** A third approach allows computer hardware changes that alter how the computer is programmed. First, we can move the program to the data, instead of the other way around. Today, data is moved in and out of the CPU for computation, but with massive data, it makes more sense to move the computation to the data, as when machine learning accesses sparse data. Second, we can leverage *approximate* and *stochastic* computing. Computers today often calculate results to a higher than required accuracy and precision. Removing this waste can both save significant power and improve computing speed, as recently demonstrated by using 16-bit floating point numbers and few-bit integers for ML training and inference, respectively. Third, as progress in general-purpose processors slows, we can build computer components called *accelerators* that provide high efficiency for specific computational phases or problems, such as ML training and inference, sensory processing, and communication mediation. Fourth, we can design systems that choreograph the use of many heterogeneous hardware resources (e.g., accelerators) to mitigate multiple bottlenecks that choke effective end-to-end solutions, such as in semi-autonomous vehicle processing where ML is but one phase. None of these approaches will speed up today's software. They will all require significant investment in both hardware *and* new software, often co-designed together. New algorithms that exploit architectural innovations will provide an even greater boost in performance. This approach has the potential for a new era of expanded computing performance beyond that provided by Level 2.

**Level 4: Non-von Neumann:** The current way we compute was first articulated by John von Neumann in 1948. But there are radically different ways to compute that may be significantly better. First, *quantum computing* uses properties of quantum mechanics to solve problems far more quickly than the von Neumann approach. Quantum computing has appropriately garnered attention, but its long-term success also requires symbiotic advances in many of the other areas we have discussed. Second, *neuromorphic computers* can leverage what is known about how the human neocortex operates. A neuromorphic computer is well suited to, for example, recognize and classify patterns in text, audio or images. It is no surprise when you think of it: the human brain is remarkably efficient at such tasks. Third, *physical computing* leverages the physics of natural processes to perform complex computational tasks. For example, an optical lens easily extracts spectral information from a light source that would be computationally expensive on traditional computers. Quantum, neuromorphic and physical computing each require significant investment in all levels of the computing stack. However, the resulting performance benefits can multiply the effects of Level 1-3 approaches.

IT can enable continued success stories across US society, but, with the slowing of Moore's Law, computing's foundations are facing technological risk that can exceed the portfolios of individual companies and industries. Pre-competitive government investment is critical to enhance this foundation of future success. This fact is clearly recognized by China and the EU, as evidenced by their increased and targeted efforts. Government investment can yield improvements that multiply across the four levels described herein to enhance US leadership across industry and society. Leading in IT, including AI support, will likely have the same lasting effects as did the Internet and personal computing.

*This white paper is part of a series of papers compiled every four years by the CCC Council and members of the computing research community to inform policymakers, community members and the public on important research opportunities in areas of national priority. The topics chosen represent areas of pressing national need spanning various subdisciplines of the computing*

*research field. The white papers attempt to portray a comprehensive picture of the computing research field detailing potential research directions, challenges and recommendations.*

*This material is based upon work supported by the National Science Foundation under Grant No. 1734706. Any opinions, findings, and conclusions or recommendations expressed in this material are those of the authors and do not necessarily reflect the views of the National Science Foundation.*

*For citation use: Conte T., Foster I., Gropp W., & Hill M. (2020) Advancing Computing's Foundation of US Industry & Society.*
*https://cra.org/ccc/resources/ccc-led-whitepapers/#2020-quadrennial-papers*